\documentclass[12pt]{article}

\begin{document}

\title{Search for Missing Charmonium States in B-meson Decays}

\author{Y. ~F. ~Gu\thanks{Email: guyf@mail.ihep.ac.cn}}

\date{}
\maketitle


\begin{center}
{\footnotesize   Department of Technical Physics, School of Physics, Peking University, 
Beijing 100871, China\\
and\\
Institute of High Energy Physics, Beijing 100039, China}
\end{center}

\vspace{0.5cm}
\begin{center} April 26, 2002 \end{center}
\vspace{1.5cm}

{\noindent \rule[0.01cm]{16cm}{0.01cm}}
{\noindent \bf {Abstract}}

The recent progress in experiments at B-factories suggests
an opportunity to search for the missing charmonium states $\eta _{c}
' (2^1S_0)$ and $h_c(1^1P_1)$. The feasibility of such a search in 
B-meson decays are discussed.

\rule[0.01cm]{16cm}{0.01cm}
\newpage
\section* {1. Introduction}

\hspace{0.5cm} Of the lowest charmonium states with radial quantum number n=1 and 2 below
the open charm threshold, two singlet states, the $\eta _{c}' (2^1S_0)$ and 
the $h_c(1^1P_1)$, are still listed by the Particle Data Group in ref.[1]
as missing or needing confirmation.

Much effort has been devoted to search for these states over past two
decades. The Crystal Ball experiment claimed the first and also the 
only observation of a candidate for the $\eta _{c} '$ in the 
inclusive photon spectrum of the $\psi '$, with a mass of
$3592\pm5$ MeV and a total width of less than 8 MeV ($95\%$ C.L.)[2].  
This result, however, has never been confirmed by other
experiments, such as E760/E835[3], BES[4], DELPHI[5], and L3[6].  No
evidence has been found for the $^1P_1$ state of charmonium in $e^+e^-$
annihilation.  The discovery of $h_c$ at $ \sqrt{s} = 3526.2\pm0.15\pm0.2
$ MeV with a width of less than 1.1 MeV ($90\%$ C.L.) once reported 
by the E760 experiment in $p\overline{p}$ annihilation [7] has
not yet been confirmed by more recent higher-statistics studies at 
the E835 experiment [8], an upgraded continuation of E760. 

It was previously deemed that the search for these two singlet states 
of charmonium poses an unusual experimental challenge since they cannot be 
resonantly produced in $e^+e^-$ annihilation nor can populated by E1 decay 
of the $\psi '$ state.  However, the unsuccessful experience at the 
Fermilab Antiproton Accumulator ring [3,8] teaches us that the reason 
may not merely due to the above difficulties in $e^+e^-$ experiments.  
In this letter, I will describe a new opportunity to search for these 
states at the B-factories and will study the feasibility of utilizing 
a few B-meson decays involving a charmonium meson to that end. 
\section* {2. Physics of the $\eta _{c} '$ state}

\hspace{0.5cm} The mass of the $\eta _{c} '$ is predicted to be somewhere
in the range between 3540-3630MeV [9].  The total width of the $\eta _{c} '$ 
can be calculated by the relation [10]
\begin {equation}
\frac{\Gamma (\eta _c ')}{\Gamma (\eta _{c})}\cong \frac{\left| \Psi
(0)\right| '^{2}}{\left| \Psi (0)\right| ^{2}}=\frac{M_{\psi '
}^{2}}{M_{\psi }^{2}}\frac{\Gamma (\psi ' \rightarrow e^{+}e^{-})}{
\Gamma (J/\psi \rightarrow e^{+}e^{-})}
\end {equation}
Plugging in the appropriate PDG numbers [1] results in $\Gamma (\eta _c ')$ to be about 7.5 MeV. 

Chao, Gu and Tuan [11] deduced a relation between the decay rates
for $\eta _{c} '$ and $\eta _{c}$ to a light hadronic final state as
\begin {equation}
\frac{B(\eta _{c} ' \rightarrow h)}{B(\eta _{c} \rightarrow h)} \cong 1
\end {equation}
and estimated partial widths for individual decay modes $\eta _{c} ' \rightarrow
\eta \pi \pi$, $\eta _{c} ' \rightarrow h_{c} \gamma$, $\eta _{c} ' 
\rightarrow J/\psi \gamma$,
and $\eta _{c} ' \rightarrow gg$ to be about 140 keV, 11 keV, 5 keV, and 4 MeV,
respectively.  It is obvious that $\eta _{c} ' \rightarrow gg$ is the dominant mode 
of the $\eta _{c} '$ decays. 
According to relation (2), $\eta _{c}'\rightarrow K \overline{K} \pi $, $\eta\pi\pi$,
$\eta'\pi\pi$, and $\rho\rho$ are expected to have the largest branching fractions.

However, the relation (2) is not necessarily to be observed for specific exclusive decays. This is a situation similar to the so-called $\rho\pi$ puzzle in $J/\psi$  and $\psi'$ decays[12].  According to the hadron helicity conservation (HHC) theorem of PQCD, the decay of the $J/\psi$ ($\psi'$) into $\rho\pi$ is first-order forbidden [13]; however, while $\rho\pi$ is the largest hadronic final state in $J/\psi$ decay, with a branching fraction 1.27\%, this final state is unobserved for $\psi'$, with a limit $<2.8\times 10^{-5}$ ($90\%$ C.L.), severely deviating from the ratio of $\psi'$ to $J/\psi$ decay rates naively predicted by PQCD[12].  Similarly, the decays of the $\eta_{c}$ in two vector mesons ($\rho\rho$, $K^{*}\overline{K}^{*}$, $\phi\phi$) and in $p\overline{p}$ are also all forbidden by HHC, and yet they are actually observed to occur with relatively large branching fractions[14].  It is not clear whether the analogous decays of the $\eta_{c}'$ obey the relation (2) or are suppressed in relation to $\eta_{c}$.  Therefore, in search for the $\eta_{c}'$ state, one has to first choose those decay modes allowed by HHC. 
\section* {3. Physics of the $h _{c}$ state}

\hspace{0.5cm} The mass of the $h _{c}$ is a subject of considerable interest as it has
implications for the Lorentz nature of confinement.  With scalar
confinement the only spin-spin force is the one-gluon-exchange contact 
interaction, and as a result the $h _{c}$ state will lie at
spin-weighted center-of-gravity of the $^{3}P_{J}$ states,
\begin {equation}
m _{c.o.g.} \equiv \frac{M _{\chi _{c0}}+3M _{\chi _{c1}}+5M _{\chi _{c2}}}{9}
\cong 3525 MeV
\end {equation}
Alternatively, with vector confinement there is a $1/r$ spin-spin force, and
the $h _{c}$ state is shifted upwards from the $m _{c.o.g.}$ by about $20 MeV/c^{2}$.  A better determination of the mass of the $h _{c}$ is thus of importance for
studies of the nature of confinement, which is one of the most interesting 
questions in nonperturbative QCD. 

It is expected that the $h _{c}$ has a small width ($< 1$ MeV) and decays
predominantly to the $\gamma \eta _{c}$ final state through an E1 transition. 
A recent
calculation [15] in both conventional PQCD and NRQCD approaches provides a
number of important decay rates for $h _{c}$.  Table 1 lists the numerical
results in both approaches. 

\begin{table}
 \caption{Results of $h_{c}$ decay rates by calculations in different approaches (where 
 $\alpha _{M} / \alpha _{E} \approx 1-3$)}
 \begin{center}
  \begin{tabular}{l|l|c}
  \hline\hline

  {\sl $h_{c}$ Decay Mode} &    {\it PQCD} &    {\it NRQCD} \\
  \hline

  {$\Gamma _{total}$} &    {$ 390 \pm 10 keV $} &   {$ 980 \pm 90 keV $} \\

  {$B(\gamma \eta _{c})$} &    {$ (85 \pm 9)\% $} &   {$ (34 \pm 3)\% $} \\

  {B(light hadrons)} &    {$ (11.5 \pm 1.8)\% $} &   {$ (54 \pm 13)\% $} \\
   
  {$B(J/\psi \pi\pi) $} &   {$ (1.8 \pm 0.2)(\alpha _{M} / \alpha _{E})\% $} & { } \\

   {$B(J/\psi \pi^{0})$} &    {$ (0.15 \pm 0.02)(\alpha _{M} / \alpha _{E})\% $} & { } \\

   {B($\gamma$ + light hadrons)} &    {$0.96\% $} &   { } \\

  {$B(K \overline{K} \pi) $} &    {$ (5.5 \pm 1.7)\% $} &
   {$ (6.5 \pm 3.5)\% $} \\

  {$B(\rho \rho) $} &    {$ (2.6 \pm 0.9)\% $} &
   {$ (2.9 \pm 1.8)\% $} \\

  {$B(\pi ^{+} \pi ^{-} K ^{+} K ^{-}) $} &    {$ (2.0 \pm 0.7)\% $} &
   {$ (2.2 \pm 1.3)\% $} \\

     \hline\hline
  \end{tabular}
 \end{center}
\end{table}

\section* {4. Search for $\eta _{c} '$ and $h _{c}$ in B meson
decays}

\hspace{0.5cm} After two years' running at the KEKB and PEPII asymmetric energy $e^+e^-$
colliders, both the Belle and BABAR detectors have collected about $30 fb^{-1}$ $\Upsilon (4S)$ data sample,
which contains over 30 million $B \overline {B}$ events[15-19].  A number 
of inclusive and exclusive B-meson decays with a charmonium in the final state have been studied and related branching fractions have been measured, as is shown in
Table 2.  It is particularly interesting to observe the factorization-forbidden decay $B \rightarrow \chi _{c0} K$, which has a rate comparable to those for 
the factorization-allowed $B \rightarrow J/\psi K$ and $B \rightarrow \chi _{c1} K$ 
decays [18], and a statistically significant inclusive $\chi_{c2}$ production in B-meson decays, which is also a factorization-forbidden process[19].  The
experimental progress at B-factories suggests a good opportunity to search
for the missing charmonia in B-meson decays.

\begin{table}
 \caption{Some branching fractions for inclusive and exclusive decays of B
mesons involving charmonium measured at B-factories.}
 \begin{center}
  \begin{tabular}{l|l|c}
  \hline\hline

  {\sl Channel} &    {\it Branching fraction $(10 ^{-3} )$} &   {\it Reference} \\
  \hline

 {$B^{+} \rightarrow \eta _{c}K^{+}$} & {$1.50 \pm 0.19 \pm 0.15 \pm 0.46$} & {[16]} \\
 {$B^{+} \rightarrow J/\psi K^{+}$} & {$1.01 \pm 0.03 \pm 0.05$} & { [17]} \\
 {$B^{+} \rightarrow \chi _{c0} K^{+}$} & {$0.60 ^{+ 0.21} _{-0.18} \pm 0.11$} & { [18]} \\
 {$B^{+} \rightarrow \chi _{c1} K^{+}$} & {$0.75 \pm 0.08 \pm 0.08$} & { [17]} \\
 {$B^{+} \rightarrow \psi 'K^{+}$} & {$0.64 \pm 0.05 \pm 0.08$} & { [17]} \\
 {$B^{+} \rightarrow J/\psi K^{*+}$} & {$ 1.37 \pm 0.09 \pm 0.11$} & { [17]} \\
 {$B^{0} \rightarrow \eta _{c} K^{0}$} & {$ 1.06 \pm 0.28 \pm 0.11 \pm 0.33$} & { [16]} \\
 {$B^{0} \rightarrow J/\psi K^{0}$} & {$ 0.83 \pm 0.04 \pm 0.05$} & { [17]} \\
 {$B^{0} \rightarrow \chi _{c1} K^{0}$} & {$ 0.54 \pm 0.14 \pm 0.11$} & { [17]} \\
 {$B^{0} \rightarrow \psi ' K^{0}$} & {$ 0.69 \pm 0.11 \pm 0.11$} & { [17]} \\
 {$B^{0} \rightarrow J/\psi K^{*0}$} & {$ 1.24 \pm 0.05 \pm 0.09$} & { [17]} \\
 {$B \rightarrow \chi _{c1} X$} & {$3.32 \pm 0.22 \pm 0.34$} & { [19]} \\
 {$B \rightarrow \chi _{c2} X$} & {$ 1.53 ^{+0.23} _{-0.28} \pm 0.27$} & { [19]} \\
  \hline\hline
  \end{tabular}
 \end{center}
\end{table}
            
In comparison with the BES experiment and the proposed CLEO-c program which take
data on the $\psi '$ resonance, the B-factory experiments will have significant
advantages in looking for the $\eta _{c} '$  and $h _{c}$.  In respect of detector
performance, both Belle and Babar are modern detectors, while the CLEO-c is similar to them but will not have any superiority over them, and the BES is modeled on MARKIII, which was built two decades ago.  In respect of statistics, Belle and Babar
will each have about $400 fb ^{-1}$ of $\Upsilon (4S)$ in a few years, corresponding to about 400 million $B \overline{B}$ events, an unprecedented huge data sample compared to both BES (14 million $\psi '$s at present in store) and CLEO-c
(50-100 million $\psi '$s tentatively planned after 2003).  

Moreover, it would be
intrinsically difficult to find an $\eta _{c} '$ signal in $\psi '$ decays when the mass of the $\eta _{c} '$ is closed to that of the $\psi '$.  As is known, the rate of $\psi ' \rightarrow \gamma \eta _{c} ' $ decays
can be determined directly from the $J/\psi \rightarrow \gamma \eta _{c}$ rate.  Assuming identical $1 ^{--}$
and $0 ^{-+}$ spatial wavefunctions, these simply scale as $k _{\gamma} ^{3}$ of the photon,
therefore
\begin {equation}
\Gamma(\psi ' \rightarrow \gamma \eta _{c} ') =
\left( \frac {(M _{\psi '} ^{2}-M _{\eta _{c} '}^{2})/2 M _{\psi '}}
{(M _{J/\psi} ^{2} - M _{\eta _{c}}^{2})/2 M _{J/\psi}}\right) ^{3} \Gamma(J/\psi \rightarrow \gamma \eta _{c})
\end {equation} 
For an $\eta _{c} '$ mass of $3630 MeV/c ^{2}$ one expects a partial width 
$\Gamma(\psi ' \rightarrow \gamma \eta _{c} ')$ of about
0.1 keV or $\Gamma(\psi ' \rightarrow \gamma \eta _{c} ') \sim 5 \times 10 ^{-4}$.  Clearly this rate falls rapidly and the photon from the M1 transition would be too soft to detect as the $\eta _{c} '$ mass approaches the $\psi '$ mass.  It is obvious
that this kind of difficulty would not occur in case of B decays into
final states containing a charmoinum meson.
\section*{5. Signal and background estimation}

\hspace{0.5cm} It is seen from Table 2 that all four
factorization-allowed decays {$B \rightarrow \eta _{c}K$}, 
 {$B \rightarrow J/\psi K$},
  {$B \rightarrow \chi _{c1} K$} and
 {$B \rightarrow \psi ' K$} have
comparable branching fractions.  One may thus expect that the branching
fraction of {$B \rightarrow \eta _{c} ' K$} is just as large as those measured values.  Taking {$B(B ^{+} \rightarrow \eta _{c} ' K ^{+})=0.64 \times 10^{-3}$} and assuming $B(\eta _{c} ' \rightarrow K _{s} ^{0} K ^{+} \pi ^{-}) \cong B(\eta _{c} \rightarrow K _{s} ^{0} K ^{+} \pi ^{-}) = (1.5 \pm 0.4) \times 10 ^{-2}$, the cascade branching fraction is 
\begin {equation}
B(B ^{+} \rightarrow \eta _{c} ' K ^{+} 
\rightarrow (K _{s} ^{0} K ^{+} \pi ^{-} +c.c.) K ^{+}
\rightarrow (\pi ^{+} \pi ^{-} K ^{+} \pi ^{-} + c.c.) K ^{+})
\cong 6.6 \times 10 ^{-6}
\end {equation}
Given the accumulated number of B mesons
$N _{B} = 1 \times 10 ^{8}$(a goal to be achieved within the year), and the signal efficiency determined for $B ^{+} \rightarrow \eta _{c} K ^{+}$, $\eta _{c} \rightarrow K _{s} ^{0} K ^{+} \pi ^{-} +c.c.$, $\epsilon \cong 10\%$[16], there will be 66 events of the $\eta _{c} '$ from $K _{s} ^{0} K ^{+} \pi ^{-} +c.c.$.

As for the $h _{c}$, the decay $B \rightarrow h _{c} K$ should be also
forbidden by the factorization just as the decay $B \rightarrow \chi _{c0}
K$ and $B \rightarrow \chi _{c2} K$.  Based purely on experimental
information, since the measured $B(B \rightarrow \chi _{c0} K)$ is
comparable with $B(B \rightarrow \chi _{c1} K)$, and $B(B \rightarrow \chi
_{c2} K)/ B(B \rightarrow \chi _{c1} K) \cong 0.5$, as is seen from Table
2, it would be rather save to assume $B(B \rightarrow h _{c} K)\cong B(B
\rightarrow \chi _{c2} K)\cong 0.4 \times 10 ^{-3}$.  If $h _{c}$ is
searched by $\gamma \eta _{c}$, which is the main exclusive decay mode of
$h _{c}$, the cascade branching fraction is
\begin {equation}
B(B ^{+} \rightarrow h_{c} K ^{+} 
\rightarrow \gamma \eta _{c} K ^{+}
\rightarrow \gamma (K _{s} ^{0}  K ^{+} \pi ^{-} + c.c.) K ^{+}
\rightarrow \gamma (\pi ^{+} \pi ^{-} K ^{+} \pi ^{-} + c.c.)K ^{+})
\cong 3.5 \times 10 ^{-6}
\end {equation}
When $N _{B} = 1 \times 10 ^{8}$, and $\epsilon = 10\%$[16], there will be
about 35 events of the $h _{c}$.

There are still room for increasing statistics with increased $N _{B}$ and by
including other channels, such as $\eta _{c} ' K ^{*+}$ or $h _{c} K ^{*+}$, and by combining
$B ^{0}/\overline {B }^{0}$ with $B ^\pm$ .  One may also carry out the search in other $\eta _{c} '$ and $h _{c}$ decay modes as well as in inclusive B decays.

The dominant source of background in most cases of exclusive B decays into two-body final states containing a certain charmonium is found to be other B decays that include charmonia in the final state[17].  For the cascade decay $B ^{+} \rightarrow \eta _{c}'K ^{+}$, $\eta _{c}' \rightarrow K _{s} ^{0} K ^{+} \pi ^{-}$, the possible competing process is $B ^{+} \rightarrow \chi_{c2}K ^{+}$, $\chi _{c2} \rightarrow K _{s} ^{0} K ^{+} \pi ^{-}$. Since there is only an upper limit set for $\chi _{c2} \rightarrow K _{s} ^{0} K ^{+} \pi ^{-}$ to date, it may not be a serious background source when one searches for $\eta _{c} '$ in B decays.  As is also shown by the measurement of the decay 
$B ^{+} \rightarrow \eta _{c} K ^{+}$ with $2.7 \times 10 ^{6}$ $B \overline{B}$ events [16], the fitted combinatorial background with the peaking background together
accounts for about one-sixth of the raw yield for the $\eta _{c} \rightarrow K _{s} ^{0} K ^{+} \pi ^{-}$ mode. Hopefully a similar scenario will happen at $\eta _{c}'$. 

At present there is not any result of data analysis which provides a basis for the
discussion of the background problem in $h _{c}$ search in B decays; also 
it seems unlikely that any decay process would compete with
$h _{c} \rightarrow \gamma \eta _{c}$ and contaminate it significantly. 
For example, $B \rightarrow \psi ' K \rightarrow \gamma \eta _{c} K$ is a decay
process with the invariant mass of $\gamma \eta _{c}$ most close to the $h _{c}$ mass, but the branching fraction of the hindered M1 transition 
$\psi ' \rightarrow \gamma \eta _{c}$ [1] is expected to
be two orders of magnitude smaller than that of $h _{c} \rightarrow \gamma \eta _{c}$ calculated either by PQCD or NRQCD[15]. 
\section* {6. Conclusion}

\hspace{0.5cm} The feasibility of searching for the $\eta _{c} '$ and $h _{c}$ states at B-factories has been discussed based on information from recent experiments and calculations.  It is shown that there is a good opportunity to observe these states through some B-meson decays before the BEPC upgrade or CLEO-c program come true.

\section* {Acknowledgements}

\hspace{0.5cm} The author gratefully acknowledges K.T. Chao, Y. Ban, Y.L. Ye and J. Ying for helpful discussions. He is indebted to Y.P. Kuang for making ref. [15] available to him before publication. He also thanks J. Ying for assistance in preparing the manuscript. Special thanks are due to Y.L. Ye for all the support during the course of this work. 

{\it Note added}. Near the completion of this work, the author learned of an e-print by Suzuki [20], containing a similar suggestion on searching for $h_{c}$.

\end{document}